\documentclass{sf2a-conf2012}
\usepackage{graphicx}
\usepackage{hyperref}
\usepackage[]{natbib}  
\usepackage[cyr]{aeguill}
\usepackage{epstopdf}

\def\BibTeX{{\rm B\kern-.05em{\sc i\kern-.025em b}\kern-.08em
    T\kern-.1667em\lower.7ex\hbox{E}\kern-.125emX}}
\bibpunct{(}{)}{;}{a}{}{,}  
\newcommand{\kms}{$\rm{km~s}^{-1}$}

\newcommand{\bl}{$B_{\ell}$}
\newcommand{\vsini}{$v\sin i$}

\begin{document}

\TitreGlobal{SF2A 2012}


\title{The Magnetism in Massive Stars project: first HARPSpol discoveries}
\thanks{Based on observations collected at the European Southern Observatory, Chile (Program ID 187.D-0917)}

\runningtitle{The Magnetism in Massive Stars project: first HARPSpol results}

\author{E. Alecian} \address{LESIA-Observatoire de Paris, CNRS, UPMC, Univ. Paris-Diderot, 5 place Jules Janssen, F-92195 Meudon Principal Cedex, France}
\author{R. Peralta$^1$}
\author{M.E. Oksala} \address{Astronomick\'y \'ustav, Akademie v\v{e}d \v{C}esk\'e republiky, Fri\v{c}ova 298, 251 65 Ond\v{r}ejov, Czech Republic}
\author{C. Neiner$^1$}
\author{the MiMeS collaboration}

\setcounter{page}{237}


\maketitle


\begin{abstract}
In the framework of the Magnetism in Massive Stars (MiMeS) project, a HARPSpol Large Program at the 3.6m-ESO telescope has recently started to collect high-resolution spectropolarimetric data of a large number of Southern massive OB stars in the field of the Galaxy and in many young clusters and associations. In this contribution, we present details of the HARPSpol survey, the first HARPSpol discoveries of magnetic fields in massive stars, and the magnetic properties of two previously known magnetic stars.
\end{abstract}



\section{The MiMeS project}

The Magnetism in Massive Stars\footnote{http://www.physics.queensu.ca/$\sim$wade/mimes} (MiMeS) project aims at understanding magnetism in massive stars and its effect on the stellar evolution, interior and close environment. The project is led by G. Wade in Canada and C. Neiner in France. The main scientific objectives are: understanding the origin of magnetic fields in massive stars, and in particular the physics of fossil fields, studying the physics of atmospheres, winds, envelopes and magnetospheres of hot stars, the rotational evolution and magnetic braking, and the evolution of magnetic OB stars and origin of neutron star magnetic fields.

The main observational part of this project is the compilation of a large number of high-resolution spectropolarimetric observations that we have been acquiring since 2008 with ESPaDOnS, Narval and HARPSpol, thanks to three Large Programmes coordinated by G. Wade, C. Neiner, and E. Alecian, respectively. The observing sample has been divided into two sub-samples: the survey component (SC) and the targeted component (TC). The SC sample is intended to search for new magnetic stars among about 400 OB stars and to establish the basic statistics of magnetic fields in massive stars (e.g. incidence of the fields in massive stars, per spectral type, per mass bins, ..., as well as fields properties as a function of mass, age, ...). This sample includes about 200 OB stars that have been observed with the International Ultraviolet Explorer (IUE), which provides us with information on the stellar wind, since the shape and strengths of some resonance lines are sensitive to the wind. The TC sample contains about 40 stars that were previously known to be magnetic, or that were part of the SC sample and have been discovered to be magnetic . We intend to acquire a large number of spectropolarimetric observations well sampled over the rotating phase of the TC stars, to study in detail their magnetic fields and circumstellar environment \citep{wade09,oksala11}.

The HARPSpol part of the MiMeS observations contains 4 TC stars, about 50 IUE OB stars to complete the ESPaDOnS and Narval samples, and about 150 OB stars in 7 clusters of different ages (from 4 to 100 Myr) to study the impact of the initial conditions of formation, and age on the fields. 

\section{First results of the HARPSpol large program}

\subsection{The Survey Component of the HARPSpol sample}

\begin{figure}[ht!]
\centering
\includegraphics[width=0.31\textwidth,clip]{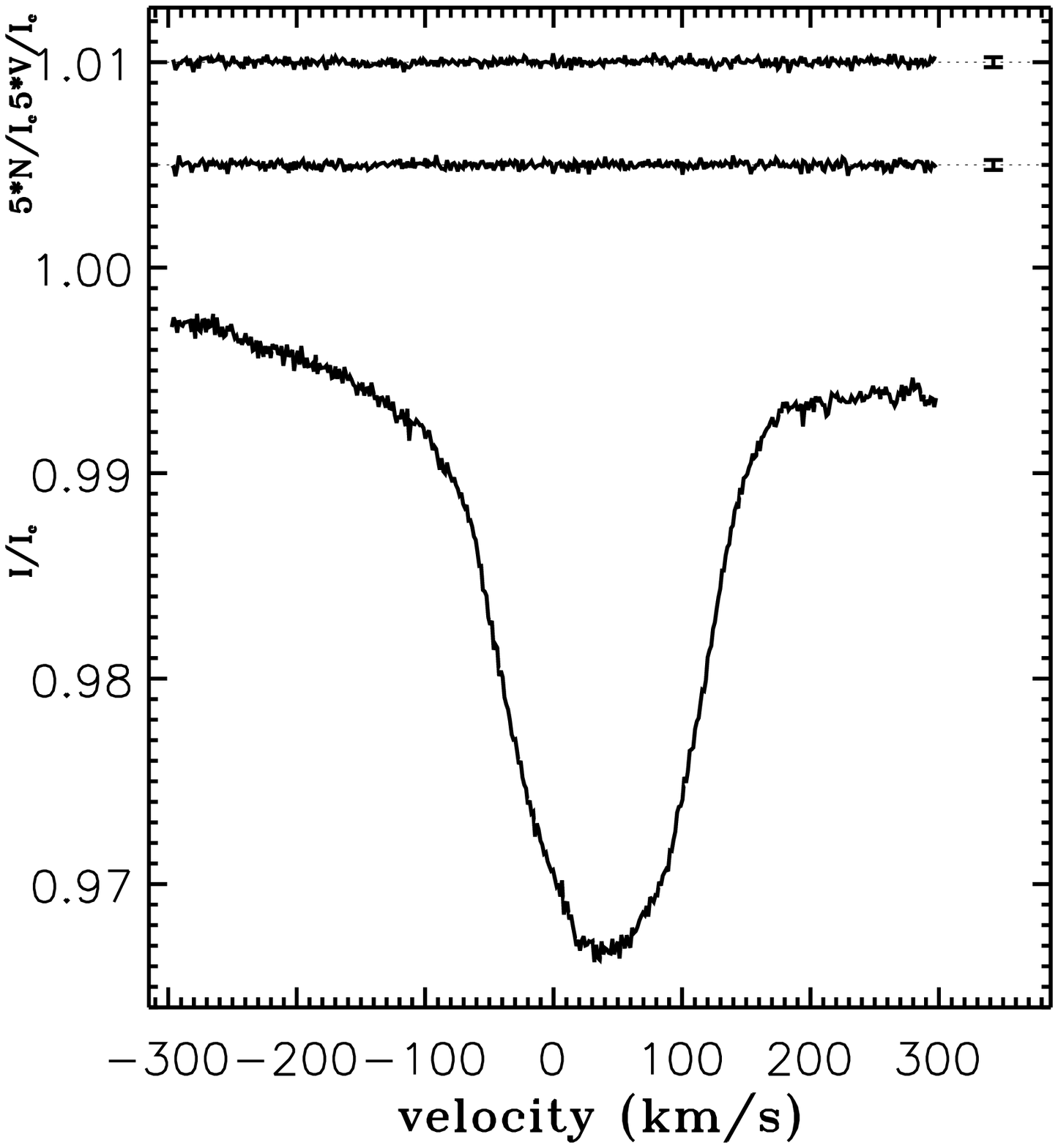}
\includegraphics[width=0.31\textwidth,clip]{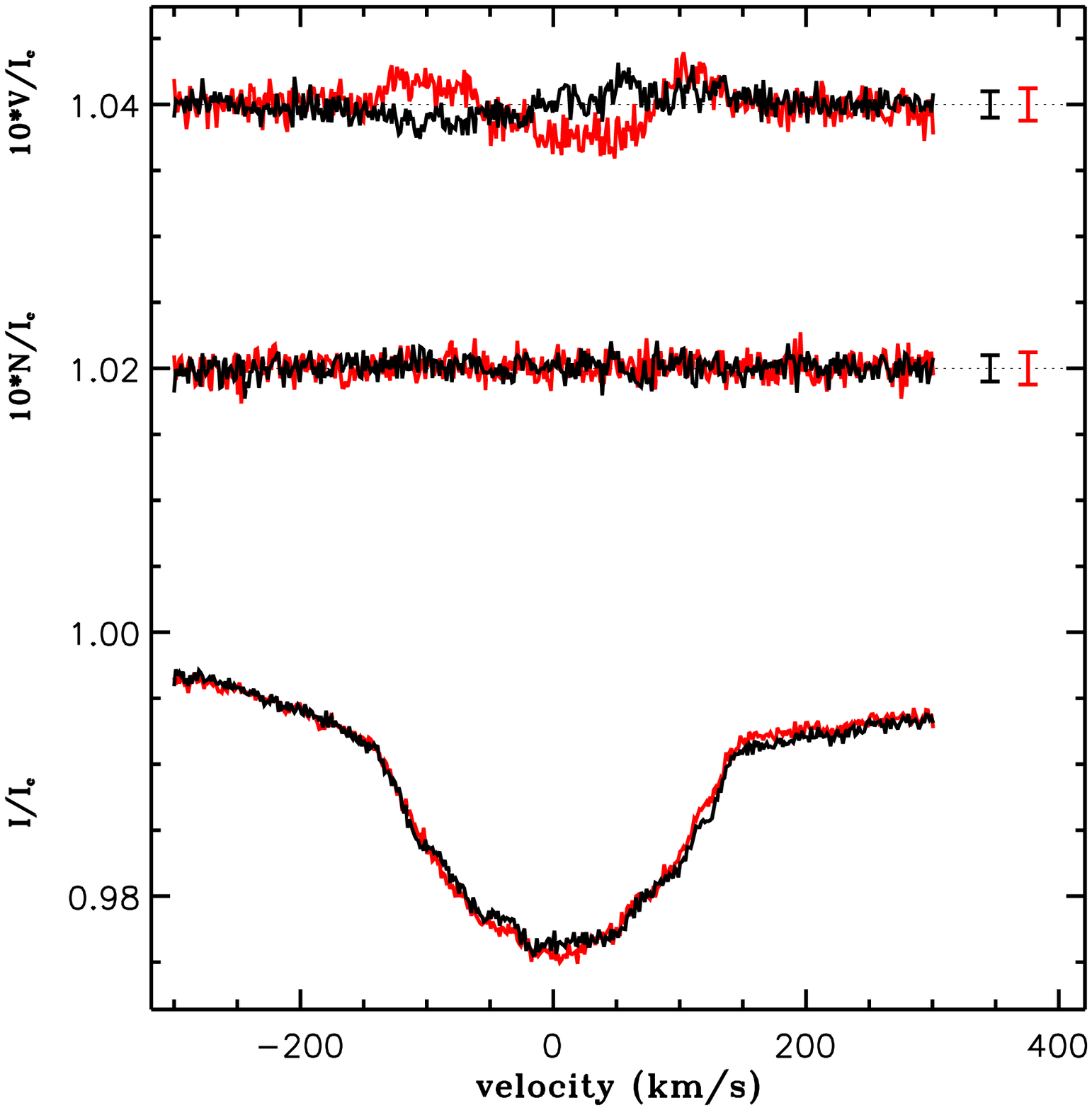}
\includegraphics[width=0.31\textwidth,clip]{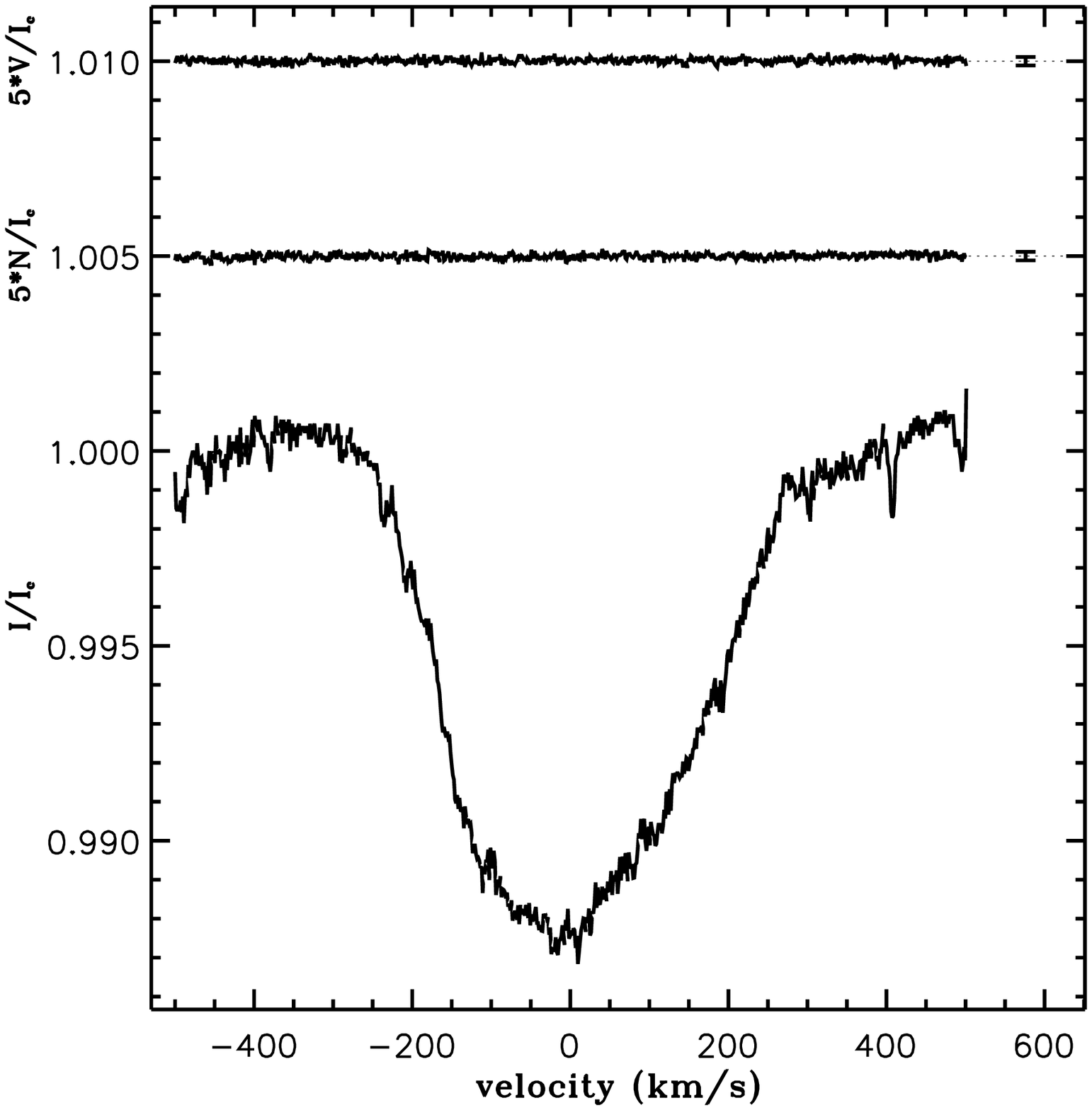}
\caption{Examples of LSD I (bottom), V (top) and diagnostic N (middle) profiles of HARPSpol targets. The observations of $\theta$ Car (left) and Achernar (right)illustrate the low upper limits on magnetic fields that we can obtain with HARPSpol. Multiple observations of the TC target HD 64740 (middle panel) are superimposed (Peralta et al. in prep.).}
\label{aleciane:fig1}
\end{figure}

\begin{figure}[t!]
\centering
\includegraphics[width=0.31\textwidth,clip]{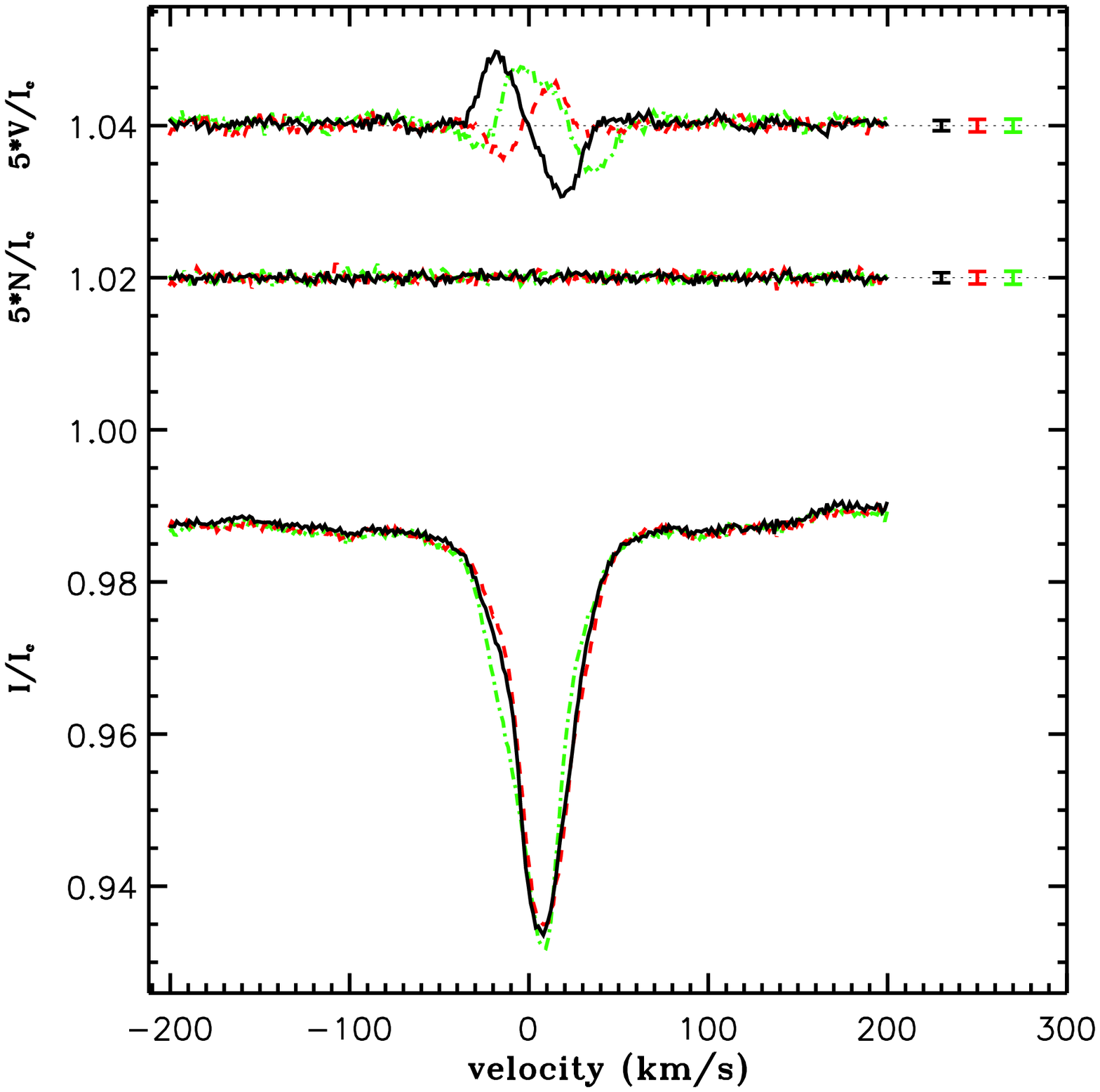}      
\includegraphics[width=0.31\textwidth,clip]{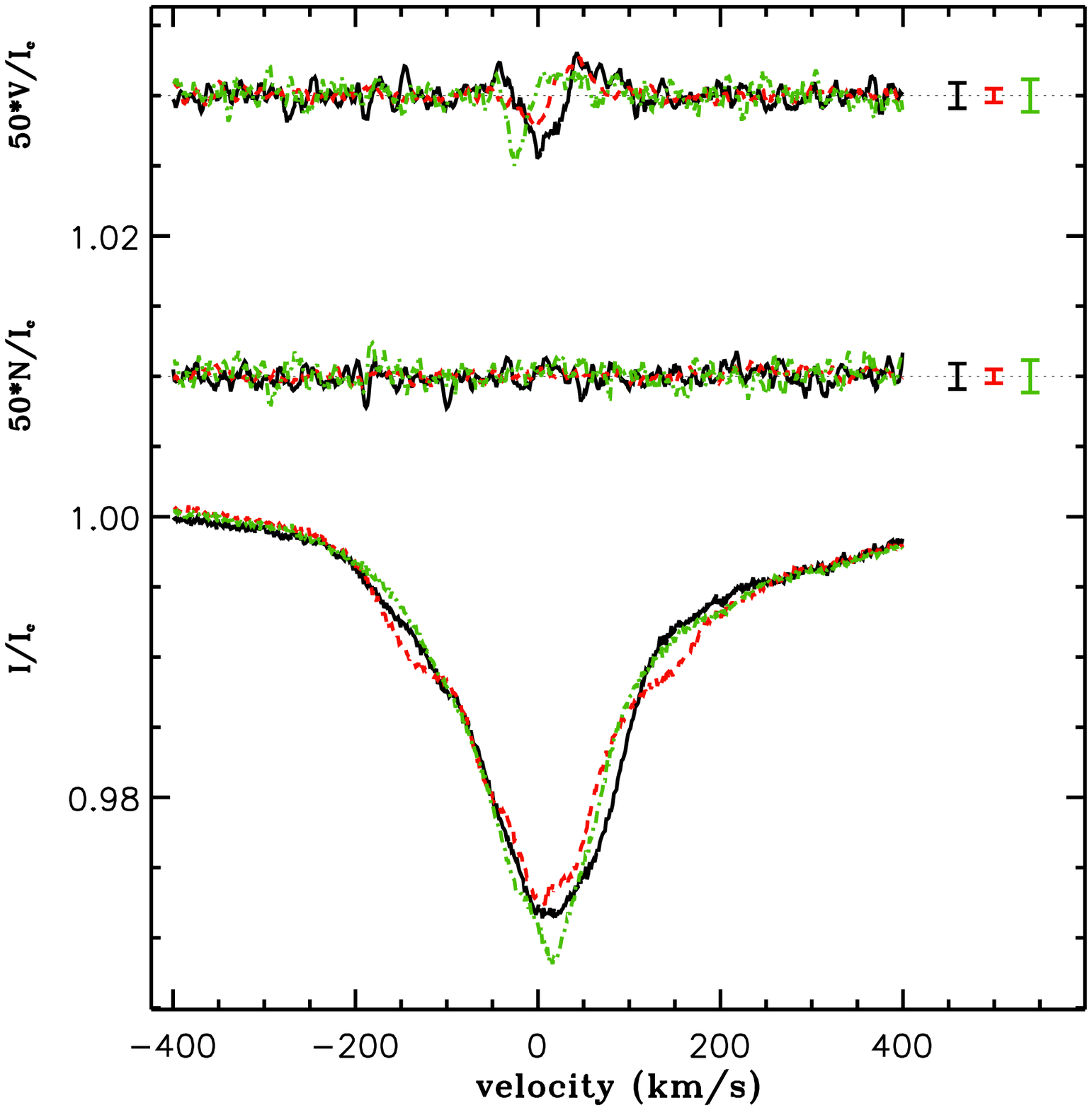}      
\includegraphics[width=0.31\textwidth,clip]{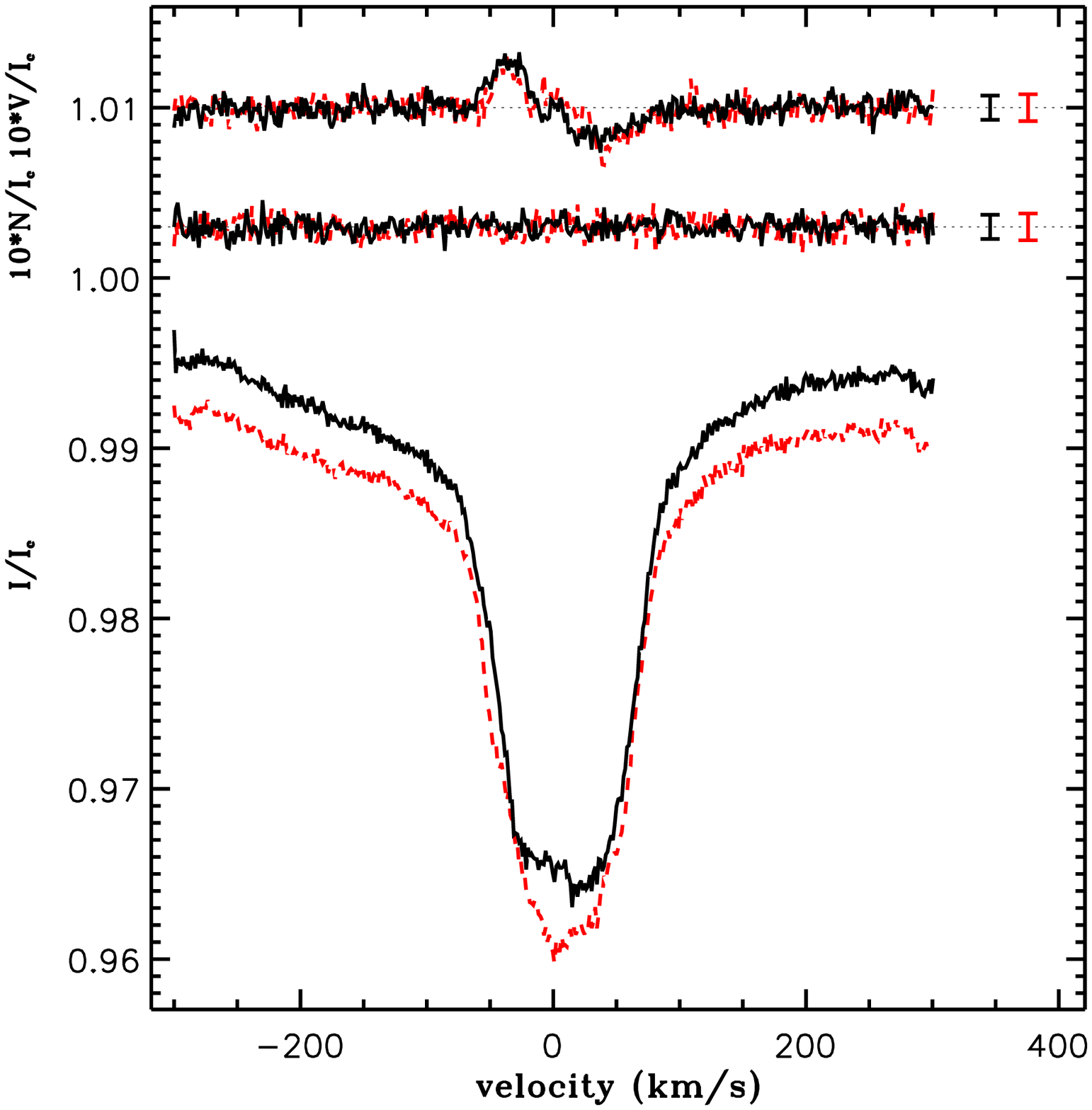}      
\caption{LSD I (bottom), V (top) and diagnostic N (middle) profiles of some of the new magnetic stars detected with HARPSpol (Alecian et al. 2011, Alecian et al. in prep.). From left to right: HD~130807, HD~122451, HD~121743. Multiple observations are superimposed for each star.}
\label{aleciane:fig2}
\end{figure}

The observations started in May 2010 and will end in February 2013. About 75\% of the observing time has been executed so far, resulting in 224 observations of 119 stars, with a resolving power of about 110\,000. We use a modified version of the REDUCE package \citep{piskunov02} to reduce the data. The Least-Square-Deconvolution \citep[LSD,][]{donati97} technique has been applied to the data to increase the signal-to-noise ratio (SNR), and hence increase the chance of magnetic detection. This technique requires the use of spectral masks \citep[see][]{donati97} that have been computed using \citet{kurucz93} ATLAS 9 atmospheric models of appropriate effective temperature and surface gravity for each star. The LSD method results in mean Stokes $I$ and $V$ profiles. A diagnostic $N$ profile is also computed in order to check that no spurious polarised signal is present in our data \citep[see][]{donati97}. Fig. \ref{aleciane:fig1} illustrates the LSD profiles obtained with HARPSpol in both magnetic and non-magnetic stars. Note that the flat $N$ profiles in all of our data indicate that no spurious signal is present, and that the Zeeman signatures in $V$, if present, are real.

\begin{figure}[t!]
\centering
\includegraphics[width=0.6\textwidth,clip]{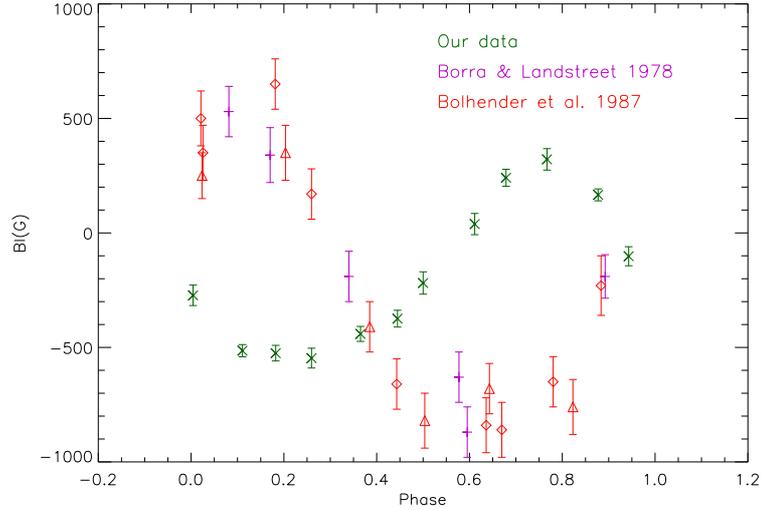}      
\caption{Longitudinal magnetic fields (\bl) vs phase of HD 64740, using the period of \citet[][BB87]{bohlender87}. Asterisks: measured on the HARPSpol spectra. Other symbols: published data of Borra \& Landstreet (1979) and BB87. The phase shift of our data is due to a lack of precision on the period. Our data show an asymmetry indicating that a quadrupolar component (in addition to the dipolar component) is present at the surface of the star (Peralta et al. in prep.).
}
\label{aleciane:fig3}
\end{figure}

The HARPSpol observations have produced 7 new magnetic stars: HD~122451, HD~130807 and HD~121743  in the Sco-Cen association, HD~66765 and HD~67621 in Vel OB2, and HD~156324 and HD~156424 in Sco~OB4 \citep[Fig. \ref{aleciane:fig2},][Alecian et al. in prep.]{alecian11}, and to confirm the magnetic field in one previously suspected magnetic star (HD 105382 of the Sco-Cen association). So far, the complete sample of OB stars has been observed for 2 out of the 7 clusters (NGC 6530 and Vel OB2). While no magnetic field has been detected among the 8 OB stars of NGC 6530, 2 magnetic stars have been detected among the 10 OB stars of Vel OB2. These results will be analysed once the whole data set of cluster stars has been acquired.

\subsection{The He-strong star HD 64740}

With HARPSPol, we have acquired 12 observations of the TC star HD 64740, well sampled over the rotation period previously determined by \citet[][$P_{\rm rot}\sim1.33026$~d]{bohlender87}. We estimated the surface averaged longitudinal magnetic fields (\bl) using the first moment technique \citep[e.g.][]{wade00}. The \bl\ are plotted with green asterisks in Fig. \ref{aleciane:fig3} together with the historic data of \citet{borra79} and \citet{bohlender87}, phased using the rotation period and the HJD at \bl\ minimum of \citet{bohlender87}. We observe a shift of about 0.5 in phase between the old and new data indicating that the rotation period is not accurate enough. Using the new and old data, we have improved the rotation period determination. The new values of the period and HJD at minimum are:
\begin{equation}
{\rm JD} (B_{\ell}\ {\rm min}) = 2455901.61503 + E1.330200(12).
\end{equation}
The He~I lines of HD 64740 are stronger than the prediction from models with a solar abundance, and vary in phase with the rotation period, confirming the He-strong nature of this star. Two patches of overabundance of helium are observed close to the magnetic poles, as usually observed in other He-strong stars.

Using the oblique rotator model of \citet{alecian08}, we modelled the Stokes $V$ profiles of HD 64740 in order to characterise its magnetic field. We find that a dipole field of 30 kG at the North pole, inclined by about 20 degrees with respect to the rotation axis, and displaced by about 0.06~R$_*$ from the centre of the star reproduce the data well (Peralta et al. in prep).

The wings of the H$\alpha$ profile of HD 64740 are superimposed with variable emission of circumstellar (CS) origin. This emission varies in phase with the rotation period and seems to be localised in two opposite clouds situated at about 3 stellar radii (Peralta et al. in prep.). These clouds are likely the result of centrifugally supported wind material trapped within the co-rotating magnetosphere, located at the intersection of the magnetic and rotational equators. This type of rigidly rotating magnetosphere (RRM) is observed in several other magnetic OB stars \citep{townsend05,oksala12,petit12}.

\subsection{The magnetic field of HD 96446}

We also obtained four observations of the TC star HD 96446 over 6 days, then, a year later, 6 observations over 10 days. HD 96446 is a well known magnetic He-strong star \citep{borra79} with a published photometric rotation period of 0.85137~d \citep{matthews91}. The data exhibits only negative Zeeman signatures, and show faint variations from one observation to the other, indicating that only the negative pole is visible. Together with a low \vsini\ ($\sim 3$~\kms), this implies that the inclination of the rotation axis is small \citep[3-15$^{\circ}$,][]{neiner12}. \citet{neiner12} have analysed the first set of HARPSPol observations and find that the field of HD 96446 is very likely dipolar with a polar strength of 5 to 10 kG. Thanks to the high spectral resolution of HARPS they are able to measure the Zeeman splitting inside the spectral lines, and find values of the field modulus compatible with a dipolar field of 5 to 10 kG. Neiner et al. conclude that while this star is very similar to $\sigma$~Ori~E, the prototype for the RRM model \citep{townsend05}, HD 96446 displays no hints of circumstellar emission in H$\alpha$. The authors propose that either the rotation period of this star should be revised significantly ($P_{\rm rot}\sim10-20$~d), or that the RRM model should be improved to include e.g. a leakage mechanism, to reproduce the observations of HD 96446.

\section{Conclusions}

HARPSpol provides us with very high quality spectropolarimetric data similar to Narval and ESPaDOnS, allowing us to reach upper limits on magnetic fields lower than 100~G on average. The very high resolution power ($\sim$110 000) allowed us to detect the Zeeman splitting inside the intensity lines of HD 96446, providing us with additional constraints on its magnetic field. Three magnetic stars among 34 have been detected in ScoCen, while 2 magnetic stars among 10 have been detected in Vel OB2. Both associations having similar ages, these first results are statistically surprising but preliminary. Serious conclusions will only be derived once the whole data set of the 7 clusters is acquired.

\begin{acknowledgements}
We wish to thank the Programme National de Physique Stellaire (PNPS) for their support.
\end{acknowledgements}

\bibliographystyle{aa}  
\bibliography{aleciane} 

\end{document}